\date{}
\documentclass[twocolumn]{revtex4}
\usepackage{graphicx,psfrag,amsmath,amssymb,amsfonts,bbm,latexsym,color,dcolumn,
epsf,graphpap}

\definecolor{red}{rgb}{1,0,0}
\definecolor{blue}{rgb}{0,0,1}
\definecolor{skyblue}{rgb}{0,0,.5}
\definecolor{green}{rgb}{0,1,0}
\definecolor{orange}{cmyk}{0,.4,1,0}

\begin{document}

\title{Environmentally induced corrections to the geometric phase in a two-level system}

\author{Fernando C. Lombardo and Paula I. Villar}

\affiliation{Departamento de F\'\i sica {\it Juan Jos\'e Giambiagi}, FCEyN UBA,
Facultad de Ciencias Exactas y Naturales, Ciudad Universitaria,
Pabell\' on I, 1428 Buenos Aires, Argentina}


\begin{abstract}
We calculate the geometric phase for different open systems (spin-boson 
and spin-spin models). We study not only how they are 
corrected by the presence of the different type of 
environments but also discuss the appearence of decoherence 
effects. These should be taken into account when planning experimental setups to study
 the geometric phase in the nonunitary regime. We propose a model 
with slow decoherence rate in which the geometric phase is still modified and 
might be measured.
\end{abstract}

\maketitle

\vspace*{0.2cm}
Since the work of Berry \cite{Berry}, the notion of geometric
phases has been shown to have important consequences for quantum
systems. Berry demonstrated that closed quantum systems could
acquire phases that are geometric in nature. He showed that,
besides the usual dynamical phase, an additional phase 
related to the geometry of the space state is generated during an
adiabatic evolution.

The existence of such a phase is also true for open quantum systems.
In particular, when a
static potential  is exerted on the
main system, the wave function of this system acquires a phase 
and hence the interference term appears multiplied by a
phase factor $e^{i \varphi}$. In an interference experiment, its effect 
on the pattern of the system is related to the phase's statistical character, 
particularly, in situations where the potential is not
static. Yet more importantly, any source of stochastic noise would create a
decaying coefficient, usually called decoherence factor $F$. For a general case, the 
phase $\varphi$ is described by means of a distribution function \cite{LomMazziVillar,Sanders}. No matter how
weak the coupling that prevents the system from being isolated, the
evolution of an open quantum system is  plagued by
nonunitary features like decoherence and dissipation. Decoherence, 
in particular, is a quantum effect whereby the system loses its ability 
to exhibit coherent behaviour and appears as soon as
the partial waves of the main system shift the environment into
states orthogonal to each other \cite{Visibility}.
Nowadays, decoherence stands as a serious obstacle in quantum
information processing. 

The geometric phase (GP) for a mixed state
under nonunitary evolution has been defined by Tong {\it et.al.}\cite{Tong} as 
\begin{eqnarray} \Phi &=&
{\rm arg}\{\sum_k \sqrt{ \varepsilon_k (0) \varepsilon_k (\tau)}
\langle\Psi_k(0)|\Psi_k(\tau)\rangle \nonumber \\
 &\times & e^{-\int_0^{\tau} dt \langle\Psi_k|
\frac{\partial}{\partial t}| {\Psi_k}\rangle}\}, \label{fasegeo}
\end{eqnarray}
where $\varepsilon_k(t)$ are the eigenvalues and
 $|\Psi_k\rangle$ the eigenstates of the reduced density matrix
$\rho_{\rm r}$ (obtained after tracing over the reservoir degrees
 of freedom). In the last definition, $\tau$ denotes a time 
after the total system completes
a cyclic evolution when it is isolated from the environment.
Taking the effect of the environment into account, the system no
longer undergoes a cyclic evolution. However, we will consider a
quasicyclic path ${\cal P}:t ~\epsilon~[0,\tau]$ with
$\tau=2 \pi/\Omega$ ($\Omega$ the system's frequency).
When the system is open, the original GP, i.e. the one that
would have been obtained if the system had been closed $\Phi^U$, is
modified. That means, in a general case, the phase is
$\Phi=\Phi^U+ \delta \Phi$, 
where $\delta \Phi$ depends on the kind of environment coupled to
the main system\cite{barnerjee}.

It is expected that GPs can be only observed in interference experiments
carried out in a time scale slow enough to ignore nonadiabatic corrections, 
but rapid enough to avoid the destruction of the interference 
pattern by decoherence \cite{Gefen}. So far, there has been no 
experimental observation of GPs for mixed states under nonunitary evolutions. The purpose 
of this short article is to study how GPs are affected by decoherence in different physical scenarios. 
The decoherence time results very important when trying to measure the GPs since for times
longer than the former the GPs, literally, disappear. In this
framework, we shall compute the GP for different models using the kinematical approach to the 
GP given by Eq.(\ref{fasegeo}), and compare the results therein obtained.
We shall start by reviewing some of our previuos results\cite{PRA}, and then  
we shall present further results concerning the environmentally induced 
corrections to the GP ($\delta \Phi$) in realistic (even experimentally feasible) models.

\vspace*{0.2cm}

{\it Purely Decohering Solvable Spin-Boson Model}. 
In this section, we shall review the basic results for an open quantum system
by presenting a model which is simple enough to be solved 
analytically\cite{PRA}. In spite of its simplicity, this model captures many of the
elements of decoherence theories and sheds some insight into the
modification of the GPs due to the presence of the environment. This model 
has been used by many
authors to model decoherence in quantum computers\cite{Ekert}
and, in particular, it is extremely relevant to the proposal 
for observing GPs in a superconducting nanocircuit \cite{Falci}.
The Hamiltonian that describes the complete evolution of the two-state system
interacting with the external environment is:
\begin{equation}
H_{\rm SB}=  \frac{1}{2} \hbar \Omega
 \sigma_z +\frac{1}{2} \sigma_z
\sum_{k} \lambda_{k} (a^{\dagger}_{k} + a_{k}) + 
\sum_{k} \hbar \omega_{k}
a_{k}^{\dag} a_{k}, \label{HSB}
\end{equation}
where the environment is described as a set of harmonic
oscillators with a linear coupling in the oscillator coordinate.
The interaction between the two-state system and the environment
is entirely represented by a Hamiltonian in which the coupling is
only through $\sigma_z$. In this particular case,
$[\sigma_z,H_{\rm int}]=0$ and the corresponding master equation is
much simplified, with no frequency renormalization and dissipation
effects. In other words, the model describes a purely decohering mechanism,
solely containing the diffusion term ${\cal D}(t)$ whose master
equation, after tracing out the environmental degrees of freedom, is given by 
(with $\hbar = 1$)  
\begin{equation}
\dot{\rho_{\rm r}} = -i \Omega [\sigma_z,\rho_{\rm r}] - {\cal D}(t)
[\sigma_z,[\sigma_z,\rho_{\rm r}]], \label{mastercap5}
\end{equation}
where
${\cal D}(s)=\int_0^s ds' \int_0^{\infty} d\omega I(\omega)
\coth\bigg(\frac{\omega}{2 k_B T}\bigg) \cos(\omega(s-s')),
$
and $I(\omega)$ is the spectral density of the environment, usually,
$I(\omega)\sim \omega^n$ up to some frequency $\Lambda$
that may be large compared to $\Omega$. In particular, the case with
$n=1$ is the ``ohmic" environment.

Then, it is easy to check that 
$\rho_{\rm r_{01}}(t)= e^{-i \Omega t -
{\cal A}(t)}\rho_{\rm r_{01}}(0)
$
is the solution for the off-diagonal terms (while
the populations remain constant), where ${\cal A}(t)= \int_0^t ds {\cal D}(t)$.
In the following, we shall call $F=\exp(-{\cal A}(t))$ the decoherence
factor.

Hence, the GP for an initial pure state of the form 
$|\Psi (0) \rangle=\cos\theta_0/2 |e \rangle + \sin\theta_0/2 |g \rangle $,
related to a quasicyclic path ${\cal P}:t ~\epsilon~[0,\tau]$  
up to first order in the dissipative constant ($\gamma_0 \propto \lambda_k^2$) is\cite{PRA} 
\begin{eqnarray}
\Phi_{\rm SB}
& \approx  & \pi (1-\cos\theta_0) 
- \frac{\gamma_0}{2} \Omega \sin^2\theta_0 
\int_0^{\tau}~dt \bigg[\frac{\partial
F(t)}{\partial \gamma_0} \bigg]\bigg|_{\gamma_0=0} \nonumber \\
& + & {\cal O}(\gamma_0^2).
\label{GPpert}
\end{eqnarray}
In the right side of last expression, we have performed 
a serial expansion in terms of $\gamma_0$. The first term corresponds to the unitary 
phase $\Phi^U$. Consequently, we see that the unitary GP 
 is corrected by a term which depends directly on the kind
of environment present \cite{PRA}. For example, for an ohmic environment
in the limit of high temperature 
$\delta \Phi_{\rm SB}^{\rm HT} = \pi^2 (\gamma_0/\Omega ) \pi k_B T \sin^2\theta_0$, while the same environment
at zero temperature modifies the unitary phase as 
$\delta \Phi_{\rm SB}^{\rm T=0}= \frac{\pi}{2} \gamma_0 (-1+\log(2 \pi
 \Lambda/\Omega))
\sin^2\theta_0$. These results can be compared with those in \cite{Sanders,TongPRA}. 
In those cases, the correction due to the environment is also proportional to $(\gamma_0/\Omega )
\sin^2(\theta_0)$ (mainly due to the simplified decoherence factor $F=\exp (-\gamma_0 t)$). However, in our model, 
these corrections enclose the main characteristic of the model of bath we are taking into account, 
which allows to evaluate the decoherence time scale properly. 

In the case of having a bosonic environment, composed by an infinite set of 
harmonic oscillators, it is not difficult to evaluate the decoherence time 
scale. This scale should be compared with the time $\tau = 2\pi/\Omega$ at which 
one expect to measure the GP. In the case of an ohmic bath in the high temperature 
limit, the decoherece time is $t_D = 1/(\gamma_0 \pi k_B T)$, which is really a 
very short time scale compared with $\tau$. In the zero temperature case, the 
decoherence time scales as $t_D \sim e^{1/\gamma_0}/\Lambda$ which, indeed, can be very 
large in the case of underdamped environments. In conclusion, one 
could expect that the GP can be only detected at very low temperature when the 
atom is mainly coupled to a bosonic field\cite{PRA}.

\vspace*{0.2cm}
{\it Spin-Spin Model}. 
 We shall study another simple solvable model in which
 the size of the environment has a relevant role. Consider a two-level
system coupled to $n$ other two-level systems\cite{Zurek}. Our main subsystem
(one qubit) interacts with the rest of the environmental spins by a bilinear 
interaction described by the interaction hamiltonian
\begin{equation}
{\cal H}_{\rm SS}=\frac{\pi}{2} \sum_{k=2}^{N} J_{1k}\sigma_z^1\sigma_z^k,
\label{HSS}
\end{equation}
where the system qubit is denoted by the superscript ``1".
This coupling is also a purely phase damping mechanism, as
in the spin-boson model mentioned above. Given a factorizable initial state
of the form $| \Phi(0) \rangle_{1}= [a |0\rangle_1 + b |1\rangle_1] 
\prod_{k=2}^n(\alpha_k |0\rangle_k + \beta_k |1\rangle_k)$,
the interacction entangles the state of the system with the environment.
This means that after the interaction, both system and environment
states are not longer factorizable. Similarly to the spin-boson model,
the density matrix will have constant populations
 (since $[\sigma_z,{\cal H}_{tot}]=0$) 
and the off-diagonal terms will be multiplied
by a decoherence factor, as $\rho_{01}^s= a b^*z(t)$ where
\begin{equation}
z(t)=\prod_{k=2}^N [ \cos(\pi J_{1k} t) + i \phi_+ \phi_- \sin(\pi J_{1k} t)],
\end{equation} where $\phi_\pm = |\alpha_k| \pm |\beta_k|$. 
Note that $z(t)$ depends on the initial conditions of the environment
only through the probabilities of finding the system in the eigenstates
of the interaction Hamiltonian $|\alpha_k|,|\beta_k|$ \cite{Zurek}.
In this case, $z(t)$ plays the role of the decoherence factor $F$ since contains
the information related to the tracing out of the spin environment degrees of freedom. In 
particular, the magnitude of $z(t)$ determines the damping of the phase information originally contained
in $\rho_{01}(0)$. In particular, when $|z(t)| \rightarrow 0$, the nonunitary evolution
and the irreversibility of the process are evident.  However, information can be
in principle recoverable for a finite system since $|z(t)| $ is at worst quasiperiodic\cite{Zurek}.
The effectiveness of the decoherence mechanism 
 is determined by the dimension of the environment.
However, in any case, if $z(t)$ is a complex function, it implies a phase shift
and an attenuation of the interference fringes, i.e. a dephasing or decoherent
process. In principle, the correction induced on the GP is the same as in Eq.(\ref{GPpert}), 
just replacing $F(t)$ by $z(t)$.

Let's take for example the particular case when the environment is composed of
only one spin ($k=2$ in Eq.(\ref{HSS})).  For the same initial state mentioned above, 
and considering $|\alpha_k|=|\beta_k|$, we obtain $z(t)= \cos(\pi J t)$ (where we set $J\equiv J_{12}$). In this
case, $z(t)$ is real and then, its only contribution is to the phase shift of
the system, while one spin environment is not effective inducing decoherence
on the system. Nevertheless, we will show that this factor induces a correction 
to the GP which is quadratic in the coupling strength with the environment. 
In such a case, if one performs a serial expansion in powers of the
coupling constant $J$, one obtains that the modification to the
unitary phase is at second order. Thus, the correction to the unitary GP is given by
\begin{equation}
 \delta\Phi_{\rm SS}^{\rm z} \approx \frac{4 \pi^4}{3 \Omega^2} 
J^2 \sin^2\theta_0 .
\label{GPss}\end{equation}
This simple result shows that correction to the unitary GP induced by the presence of 
this environment can be, in principle, detected in an interference experiment, 
without the constraint imposed by the decoherence time scale. At zero-order, the unitary 
GP is the same as in Eq.(\ref{GPpert}) $\Phi^{U} = \pi (1 - \cos\theta_0)$.

\vspace*{0.2cm}

{\it Hierarchical Qubit-Qubit Decoherence Model}. 
Herein, we shall compute the GP's correction for a model
very similar to the above described spin-spin one. This scenario
has the particular feature that it can be implemented to simulate
quantum decoherence \cite{Teklemariam}. In this case, the environment
is also limited to only one spin (qubit). However, through the strategy
of randomly redressing the phase of the environment qubits during 
the interaction with the system, it is possible to simulate a much
larger environment. Therefore, the result must be averaged over many realizations
of this evolution.  The dimension of the Hilbert space can not be 
larger than $N^2$, where $N$ is the dimension of the local
main system. To remove the information from the finite
quantum environment, a classical stochastic field is included.
Basically, the technique consists of applying classical kicks to
the environment qubits, and then averaging over the realizations of
this stochastic noise. This has the effect of scrambling 
the system information after it has been stored in the quantum
environment through the coupling interaction.

We shall consider the
evolution of this system subject to a sequence of kicks that only
affect the environment qubit. Every kick is generated by a
transverse magnetic field whose effect is to rotate the
environment qubit around the $y$ axis by an angle $\epsilon$
included randomly in the interval $(-\alpha,\alpha)$. In this
case, the reduced density matrix is similar to the above models,
but for a different decoherence factor $F$. The off-diagonal terms are
$\rho_{{\rm r}_{ij}}=a b^*f_{ij}$, where $f_{ij}$ carries all the
information about the effect of the environment qubit on the
system qubit. It is obtained after tracing out the environment
degree of freedom and averaging over the many realizations of the
external magnetic field \cite{Teklemariam}. In the case that there
are no kicks, i.e. $\alpha=0$, and $f_{12}=\cos(\pi J t)-i p_z
\sin(\pi J t)$, which agrees with the spin-spin model
described above ($p_z$ is the initial polarization of the environment qubit). In this case there is 
no decoherence and the GP-correction is given by Eq.(\ref{GPss}). 
The decoherence factor is independent of the kicking rate (no kicks 
in this limit), and the system qubit rotates independently of the 
environment qubit. 

If one allows a complete ramdomization, i.e. the kick angles $\epsilon_j$ 
may vary over the entire interval between $0$ and $2\pi$, the 
decoherence factor can be approximated, in the limit of faster kicks, by\cite{Teklemariam}
$f_{01}(\Gamma, t) \approx e^{- \frac{\pi^2J^2 t}{2\Gamma}}
- i p_z \sin(\frac{\pi J}{\Gamma}) e^{- \frac{\pi^2 J^2 t}{2\Gamma}}$
,
where $\Gamma$ is the kick rate. Using this expression, one can evaluate the 
correction induced on the GP ($\delta\Phi_{\rm SS}^{\rm cr}$) (for the particular 
case $p_z = 0$) as 

\begin{equation}
\delta\Phi_{\rm SS}^{\rm cr} \approx \frac{\pi^4}{2 \Gamma \Omega} J^2 \sin^2\theta_0 .\end{equation} In this situation, the 
decoherence time is given by $t_D = 2\Gamma/(J^2)$ which is larger than $\tau$, making the decoherence process
 negligible if trying to measure these corrections to the GP.

Finally, we shall consider the case of small angles, since
it is the regime used by simulations and 
also for decoherence experiments (usually with $\alpha=\pi/20$). In such a
case, it is possible to estimate the decoherence factor as
$f_{12}=e^{-\Gamma t \epsilon}(1+ \epsilon/2)[\cos(\pi J t)-i
p_z \sin(\pi J t)]$, where $\epsilon=2/3 \alpha^2$ is a small number ($\epsilon \approx 0.016$ for the given 
experimentally accesible value of $\alpha$ mentioned above). This decoherence factor determines a very large dephasing scale: $t_D = 1/(\Gamma\epsilon)$. In this case, we can also evaluate the 
environmentally induced correction to the GP (up to sencond order in the coupling with 
the environment and also for small $\epsilon$, and $p_z=0$) $\delta\Phi_{\rm SS}^{\rm sa}$ as  
\begin{equation}
 \delta\Phi_{\rm SS}^{\rm sa} \approx \frac{\pi}{\Omega} \sin^2\theta_0 \left[\left(\pi \Gamma - \frac{\Omega}{2}\right) \epsilon + \frac{2}{3}\frac{\pi^4}{\Omega} J^2 \right].\label{correctionsa}
\end{equation} This correction to the GP has a term independent of the coupling constant with 
the environment $J$, which in this limit is linear with $\epsilon$, the small angle that 
is rotated due to the kicks. It is worthly noticing that in the limit of $\epsilon \rightarrow 0$, Eq.(\ref{correctionsa}) 
coincides with the result given by the Zurek's model. 

Even though this is a very simple quantum open system model, it is of great 
interest due to the fact that this scheme enables simulation of the quantum 
decoherence that usually appears for larger environments. As we have mentioned, 
one qubit as environment is not enough to produce 
decoherence on the system qubit in the Zurek's model. However, in the present case, 
the phase damping is induced 
by a sequence of kicks that affect only the environment qubit, generated by a magnetic field
 that rotates the environment spin by an angle $\epsilon$. We believe that this 
practical implementation could be suitable for measuring of the complete GP in the 
case of a nonunitary evolution.

This work was supported by UBA, CONICET, and ANPCyT, Argentina.

\end{document}